\documentclass[preprint,12pt]{elsarticle}

\usepackage{amssymb}

\journal{Nuclear Instruments and Methods in Physics Research Section A}

\begin{document}

\begin{frontmatter}



\title{Development of liquid $^{3}$He target for experimental studies of antikaon-nucleon interaction at J-PARC}


\author[a,b]{M. Iio \corref{cor1}}
\ead{masami.iio@kek.jp}

\author[b]{S. Ishimoto}
\author[c]{M. Sato}
\author[d]{S. Enomoto}
\author[c]{T. Hashimoto}
\author[b]{S. Suzuki}
\author[a,e]{M. Iwasaki}
\author[c]{R.S. Hayano}

\cortext[cor1]{Corresponding author}

\address[a]{RIKEN Nishina Center, RIKEN, Saitama 351-0198, Japan}
\address[b]{High Energy Accelerator Research Organization (KEK), Ibaraki 305-0801, Japan}
\address[c]{Department of Physics, The University of Tokyo, Tokyo 113-0033, Japan}
\address[d]{Department of Physics, Osaka University, Osaka 560-0043, Japan}
\address[e]{Department of Physics, Tokyo Institute of Technology, Tokyo 152-8551, Japan}

\begin{abstract}
A liquid $^{3}$He target system was developed for experimental studies of kaonic atoms and kaonic nuclei at J-PARC. $^{3}$He gas is liquefied in a heat exchanger cooled below 3.2 K by decompression of liquid $^{4}$He. To maintain a large acceptance of the cylindrical detector system for decay particles of kaonic nuclei, efficient heat transport between the separate target cell and the main unit is realized using circulation of liquid $^{3}$He. To minimize the amount of material, a vacuum vessel containing a carbon fiber reinforced plastic cylinder having an inside diameter of 150 mm and a thickness of 1 mm was produced. A target cell made of pure beryllium and beryllium-aluminum alloy was developed not only to minimize the amount of material but to obtain also high x-ray transmission. During a cooling test, the target cell was kept at 1.3 K at a pressure of 33 mbar. The total estimated heat load to the components including the target cell and heat exchanger cooled by liquid $^{4}$He decompression, was 0.21 W, and the liquid $^{4}$He consumption rate was 50 L/day.
\end{abstract}


\end{frontmatter}



\section{Introduction}
\label{sec:1}

In recent years, the study of kaonic atoms and kaonic nuclei has become a hot topic because of the possibility that a high-density nuclear medium is created around antikaons. Experimental data from kaonic hydrogen x-ray spectroscopy at KEK\cite{iwa97} have confirmed the attractive nature of the $\bar{K}$p interaction. This fact allows us to assume that $\Lambda$(1405) is a $\bar{K}$p bound state due to the strong interaction. On the basis of this assumption, Akaishi and Yamazaki predicted meta-stable kaon bound state formation in light nuclei using a variational calculation\cite{aka02, yam02}. Their prediction has triggered many experimental searches\cite{ang05, ben07, suz07, kis07, sat08, yim10, yam10} as well as theoretical studies\cite{she07, ike07, iva05, nis08, wyc09, dot09, yam09, ari08}. Various candidates for the deeply bound kaonic nuclear state were reported by several international collaborative experiments, but it is difficult to understand all data consistently. 

To clarify the situation, a next-generation experimental search for kaonic nuclei by the in-flight $^{3}$He($K^{-}, n$) reaction (E15)\cite{iwa06}  is performed at the hadron facility of Japan Proton Accelerator Research Complex (J-PARC)\cite{tan10}. The advantage of the E15 experiment is that it simultaneously performs exclusive measurement by a missing mass study using the primary neutron and invariant mass spectroscopy via the expected decay ($K^{-}pp \rightarrow \Lambda p \rightarrow p\pi^{-}p$).

On the other hand, more precise experimental data on kaonic atoms are necessary to completely understand the $\bar{K}$N interaction. They also supply extremely important information about the existence of kaonic nuclei. Therefore, precision spectroscopy of kaoinc $^{3}$He atom x-rays (E17) is performed at J-PARC\cite{hay06}. The purpose is to measure the 2$p$-level shift and the width of kaoinc $^{3}$He with a precision of a few eV.

Because these experiments consider kaonic atoms and kaonic nuclei, their physicals motivations are closely related. A liquid $^{3}$He target is an essential component of the experimental setup for both of them. This paper describes in detail the development and evaluation of the liquid $^{3}$He target system for the J-PARC E15/E17 experiments.
.

\section{Structure of the liquid $^{3}$He target system}
\label{sec:2}
The present cryogenic system for the liquid $^{3}$He target is shown in figure 1. The characteristics and experimental requirements of the target system are as follows:

\begin{figure}[htb]
\centering
\includegraphics[width=0.75 \columnwidth,angle=0]{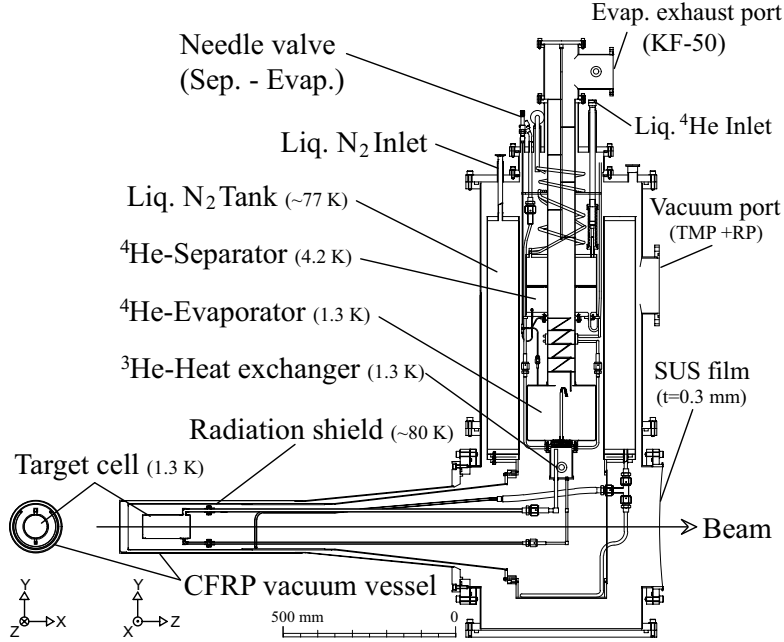}
\caption{Schematic side view of cryogenic system for liquid $^{3}$He target.}
\label{fig.1}       
\end{figure}

\begin{enumerate}

\item $Cryogenic\ system.$ 
To liquefy $^{3}$He gas, which boils at 3.2 K at 1 atm, the target system is equipped with a cryostat developed for KEK-PS E471/E549 experiments\cite{sat09} that uses liquid $^{4}$He decompression. The liquid $^{4}$He is transferred from an outside dewar vessel to a separator by a pumping method and from the separator to the evaporator via a needle valve that precisely controls the flow of liquid $^{4}$He. In the one-shot operation mode, in which the liquid $^{4}$He supply to the evaporator is stopped periodically after filling, the achieving temperature of the cryostat achieves a temperature of 1.3 K at a heat load of 0.16 W using a pump at a pumping speed of 120 m$^{3}$/h. A heat exchanger for liquefaction of the $^{3}$He gas is placed at the bottom hole of the evaporator; and its temperature is controlled by the pressure in the evaporator. The heat exchanger is described in detail in the next section. To reduce the heat load by radiation at room temperature, the cryostat is surrounded by a thermal shield cooled by liquid N$_{2}$.

\item $Siphon\ method.$ 
In the E15 experimental setup, the target cell is placed at the center of the cylindrical detector system. To maintain a large acceptance in not only forward detectors but also the cylindrical detector system, the target cell is separated from the vertical-type cryostat. $^{3}$He condensed in the heat exchanger moves to the target cell $\sim$1.2 m away through a lower pipe by gravity. The $^{3}$He vapor returns to the heat exchanger through the cell's upper pipe. Finally, the heat load at the cell is transported to the heat exchanger by circulation of liquid $^{3}$He itself.

\item $Minimum\ amount\ of\ material.$ 
The low-temperature target system includes many structures, including the target cell, the radiation shield, and the vacuum vessel between the target material and the detectors. The amount of material in these structures must be minimized in order to transmit low-energy particles and decrease background events.

\item $Stable\ maintenance\ of\ liquid\ ^{3}He.$ 
Because $^{3}$He gas is scarce, a gas handling system was produced to reuse it many times during development and the experimental period. It makes it possible to store, inject, and recover $^{3}$He gas without allowing it to escape to the atmosphere owing to damage along the route by sudden changes in the pressure and stress. The gas circuit, including the gas handling system, is shown in figure 2. $^{3}$He gas (400 L) at atmospheric pressure is stored in two 200 L tanks mounted on the system. During most of the operation period, all the valves placed between the heat exchanger and both gas tanks are open, and they reach pressure equilibrium with vapor pressure depending on the temperature in the heat exchanger. The gas system is oil-free because it uses a leak-tight scroll pump for recovery. As a safety measure, $^{3}$He gas that leaks in the insulation vacuum is collected once in a 1000 L balloon in the gas system. Afterwards,$^{3}$He gas laced with air and oil is returned to the tanks after purification by a liquid N$_{2}$ refiner.

\end{enumerate}

\begin{figure}[htb]
\centering
\includegraphics[width=1 \columnwidth,angle=0]{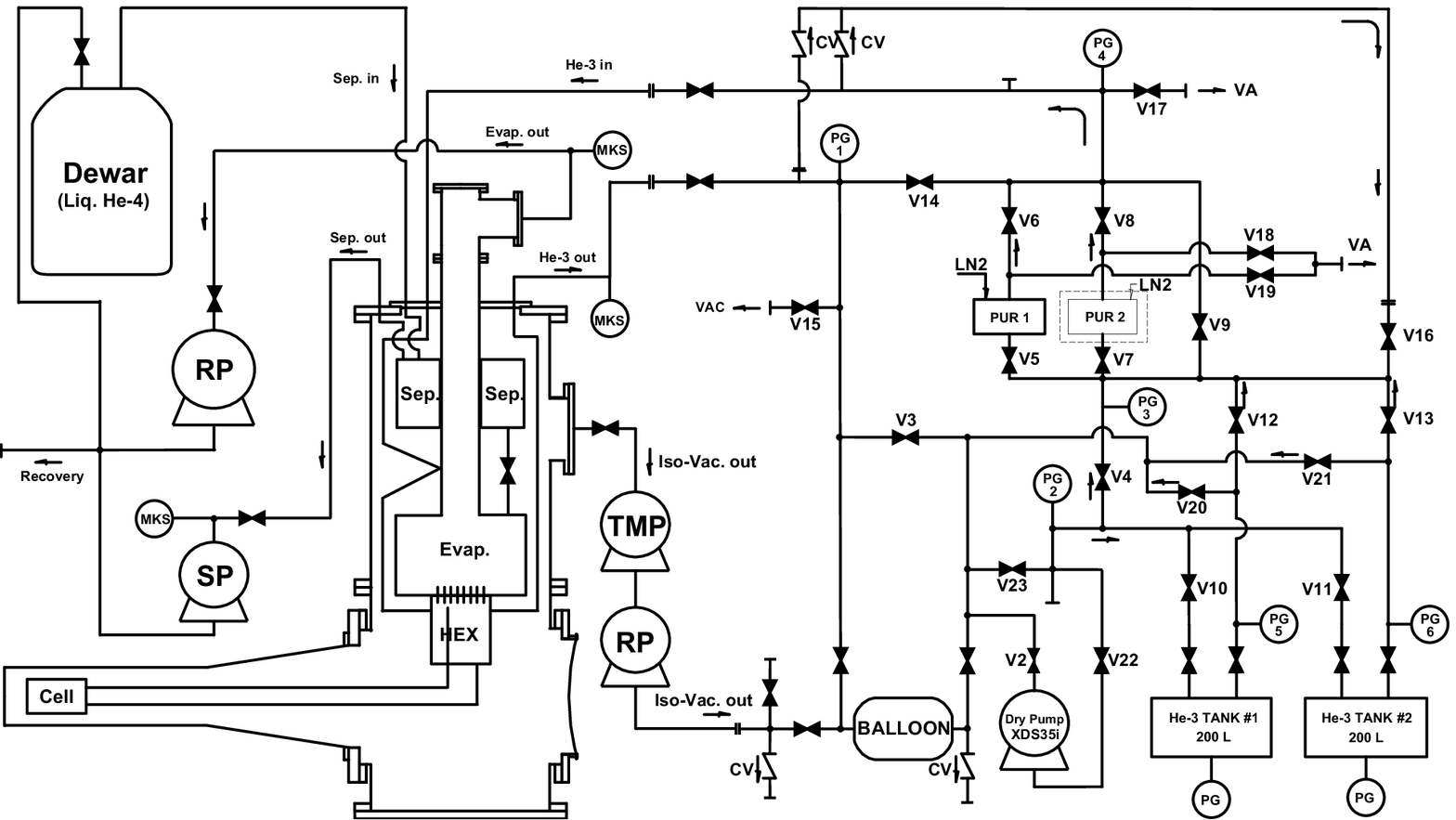}
\caption{Gas circuit of the liquid $^{3}$He target system.}
\label{fig.2}       
\end{figure}

\section{Development of the main components}
\label{sec:3}

\subsection{$^{3}$He heat exchanger}
\label{subsec:3.1}
A schematic view of the evaporator and heat exchanger is shown in figure 3. The Cu fin structures, which have a width of 0.5 mm and distance of 0.5 mm are attached to top of the heat exchanger by silver alloy brazing. High heat-transfer efficiency is obtained by direct contact between the fins and the liquid $^{4}$He in the evaporator. $^{3}$He gas is transported from the two tanks of the gas handling system through an inlet pipe thermally anchored on top of the separator and an exhaust duct on the evaporator to the heat exchanger. Liquefied $^{3}$He is transported through a pipe of 1/4" in diameter from the bottom of the heat exchanger to the target cell $\sim$1.2 m away by gravity. The pipe is removed by a crank structure from the flight path of beam particles and  forward neutrons from the ($K^{-}, n$) reaction. The $^{3}$He vapor that returns from the cell through a pipe 3/8" in diameter is recondensed in the heat exchanger by a siphon method. Finally, the heat load to the cell is transported to the heat exchanger by circulation of liquid $^{3}$He itself. The internal temperature of the heat exchanger and the target cell are each measured by a platinum-cobalt resistance temperature detector (Chino, R800-6)\cite{chino} and calibrated ruthenium dioxide resistance temperature detector (Scientific Instruments, RO600) with an accuracy of $\pm$ 0.010 K below 1.5 K\cite{scins}. The cables of the thermometer placed in the cell and the heat exchanger are connected to the hermetic seal pins at the feed-through port (KF-16) of the heat exchanger. During operation, an outlet pipe line isolated from the tanks is used to monitor the pressure in the heat exchanger using an absolute pressure transducer (MKS Instruments, Baratron 627B)\cite{barat}. The results of the temperature and pressure measurement are described in the next section.

\begin{figure}[htb]
\centering
\includegraphics[width=0.4 \columnwidth,angle=0]{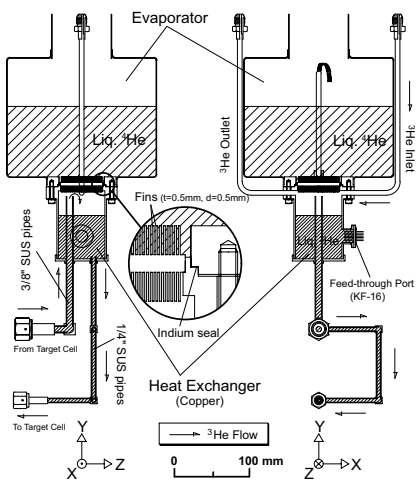}
\caption{Schematic view of evaporator and heat exchanger.}
\label{fig.3}       
\end{figure}

\subsection{Target cell}
\label{subsec:3.2}
The target cell must be designed to be pressure tight and leak-tight under a low-temperature environment in order to hold liquid $^{3}$He stably. although the cell is maintained at a pressure of less than 0.1 MPa during operation, it is designed for safe operation at an internal pressure of more than 0.3 MPa. The allowable leak rate of the cell is less than order of 10$^{-10}$ Pa$\cdot$m$^{3}$/sec by measurements using superfluid $^{4}$He. If only those requirements are considered, heavy metal such as copper, iron, and stainless steel is advantageous to producing the cell. However, the E15 experiment requires the amount of materials to be minimized in order to increase the transmission of low-energy particles and decrease the number of background events. In addition, because the target system is also used for precision x-ray spectroscopy, the experimental requirements were expanded to include the adoption of materials that provide high x-ray transmission (K$^{-}$-$^{3}$He 3$d$$\rightarrow$2$p$: 6.2 keV) and the exclusion of elements that emit background x-rays (O, Fe, Cu, and Ti) for use inside the solid angle of the x-ray detector. It was quite challenging to satisfy all the requirements. By trial and error including the high-polymer material as well as metal, beryllium and aluminum were adopted as the principal materials of the target cell.

A schematic view of the target cell is shown in figure 4. Because the volume of the cell is 0.48 L, 269 L of $^{3}$He gas is necessary to fill it with liquid at a temperature of 1.3 K. The cell consists of a beam window, a cylinder, a ring frame, and end cap. The beam window and end cap are made from an alloy of 62\% pure beryllium and 38\% pure aluminum (Brush Wellman, AlBeMet AM162)\cite{brush}. The window and cap are both 0.6 mm thick in the beam direction. Because x-ray detectors are installed around the cell perpendicular to the beam direction, a cylinder having 68 mm inside diameter, 0.3 mm thick, and 100 mm long was produced of  99.40\% pure beryllium (Brush Wellman, PF-60)\cite{brush}. The transmissivity of Cu-K$\alpha$ x-rays (8.014 keV) of 0.305 mm thick PF-60 is 89.5\% according to the catalog specifications. The 1/4" pipe is attached to the lower part and the 3/8" pipe is attached to the upper part of an aluminum ring frame. The temperature in the cell is monitored by the ruthenium oxide resistance temperature detector\cite{scins} in the cell's upper pipe. All components were joined by electron beam welding.

A cooling test for performance evaluation demonstrated that liquid $^{4}$He at 1.3 K was maintained stably in the cell for few days. The leak rate of $^{4}$He was stable at less than order of 10$^{-10}$ Pa$\cdot$m$^{3}$/sec during the entire test period. Thus, good leak-tight performance was confirmed without super-leaks.

\begin{figure}[htb]
\centering
\includegraphics[width=0.4 \columnwidth,angle=0]{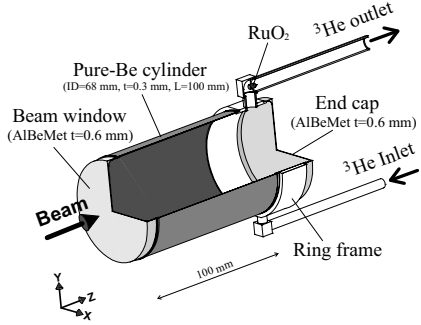}
\caption{Schematic view of target cell.}
\label{fig.4}       
\end{figure}

\subsection{Radiation shield and vacuum vessel}
\label{subsec:3.3}
In the E15 experimental setup, the detectors are located not only on the beam line but also on the circumference of the target; therefore, the amount of material in the radiation shield and vacuum vessel are also minimized. The base of the shield and the vessel were produced in a tapered form to secure an effective area for the downstream detectors. Figure 5 shows a side view of the radiation shield and vacuum vessel around the target.

\begin{figure}[htb]
\centering
\includegraphics[width=0.8 \columnwidth,angle=0]{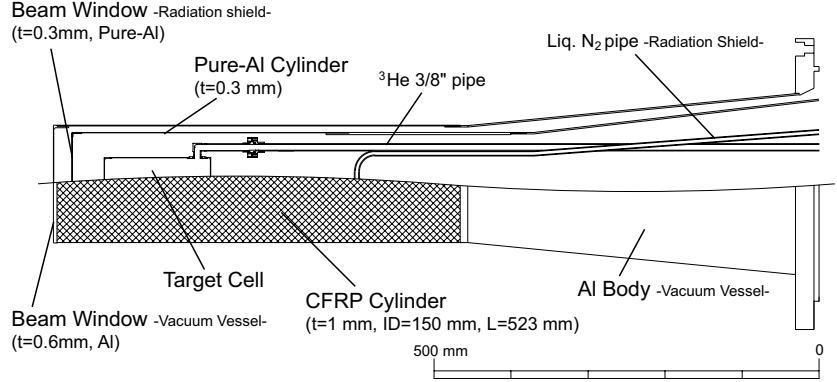}
\caption{Side view of radiation shield and vacuum vessel around the target cell.}
\label{fig.5}       
\end{figure}

The shield for reducing the heat load caused by radiation to the target cell and other components having temperatures less than 4.2 K is cooled to about 80 K by liquid N$_{2}$. The tip of the shield around the cell is made of 0.3 mm thick aluminum to reduce quantity of material. The tip, including the cap, is cooled using the good thermal conductivity of highly-pure aluminum.

Carbon fiber reinforced plastic (CFRP) was adopted for the cylinder around the target region of the vacuum vessel because of its light weight and high strength. The cylinder must have 150 mm inside diameter, more than 300 mm long, and less than 1 mm thick. In addition, for safety reasons, it must not be destroyed under an external pressure of 0.3 MPa. However, it is difficult to design structures using CFRP because many parameters must be considered, such as the strength of the carbon fiber, strength of the plastic, direction of the fiber, and method of knitting the fiber. Therefore, by using past development experience\cite{sat09}, we determined the effective Young's modulus at which the safety factor became 1 when an external pressure of 0.1 MPa acted on a prepreg laminating cylinder having 300 mm inside diameter, 250 mm long, and 0.9 mm thick.

Two CFRP cylinders were produced by Sankyo Manufacturing Co, Ltd.\cite{sankyo}. The specifications of the prepregs used for production, which were made by Mitsubishi Plastics, Inc.\cite{mitsu}, are given in table 1. TS and EM represent the tensile strength and elastic modulus of the fiber direction, respectively. For the first cylinder, three layers of prepreg with higher rigidity, called 80tonUD, were laminated in the circumferential direction (90$^{\circ}$) and one layer  was laminated in the longitudinal direction (0$^{\circ}$). Because 80tonUD has quite high rigidity, there was fear that the fiber is broken in the winding process. Therefore, for the other cylinder, a three laminated structure (two layers at 90$^{\circ}$, one layer at 0$^{\circ}$) made of 60tonUD type prepreg, with slightly lower rigidity, was adopted. Both cylinders were 150 mm in inside diameter, 523 mm long, and 1 mm thick. The design values of the safety factor for withstanding pressure were 3.8 and 3.3, respectively. The cylinders were glued to a beam window 0.6 mm thick and a tapered base made of aluminum with STYCAST1266. The safety of both vacuum vessels was confirmed by several pressure tests under a differential pressure of 0.1 MPa.

\begin{table}
\caption{Specifications of prepregs}
\label{SchmidtPL_tab:1} 
\begin{center}
\begin{tabular}{lcccccl}
\hline
Type & Metal(fiber/resin) & TS & EM & Thickness \\ 
\hline
80tonUD & K63A12/C333E & 1200 MPa & 460 GPa & 0.213 mm \\
60tonUD & K63712/C333E & 1100 MPa & 390 GPa & 0.292 mm \\
\hline
\end{tabular}
\end{center}
\end{table}

\section{Performance evaluation}
\label{sec:4}
Many cooling tests for performance evaluation were performed at KEK and J-PARC. Figure 6 shows the typical condensation periods of $^{3}$He gas by liquid $^{4}$He cooling after precooling by liquid N$_{2}$. The upper part of the figure shows the decrease in the amount of $^{3}$He gas remaining in the tanks according to pressure monitoring. From 2 to 3 h, liquefaction of the gas at a condensation speed of about 3.2 L/min appears. Subsequently, gas condensation stops at an equilibrium vapor pressure of about 0.3 atm at 2.3 K and the pressure in the tanks. To allow the remaining $^{3}$He gas to liquefy after storage with liquid $^{4}$He in the evaporator, the temperature in the heat exchanger was reduced to stop the flow of liquid $^{4}$He from the separator to the evaporator. A needle valve was closed at the time indicated by the dotted line in figure 6. $^{3}$He gas (380 L, 0.95 atm) was ultimately condensed. Furthermore, the temperature difference between the target cell and the heat exchanger disappeared at (A) in figure 6. Because the temperature sensor was installed in the top of the target cell, this indicates that the cell became filled with liquid $^{3}$He.

\begin{figure}[htb]
\centering
\includegraphics[width=0.4 \columnwidth,angle=0]{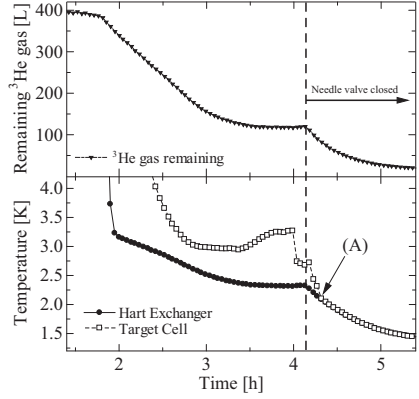}
\caption{Temperature distribution inside the $^{3}$He system and behavior of remaining amount of $^{3}$He gas in the outside tanks.}
\label{fig.6}       
\end{figure}

figure 7(a) shows a typical temperature achieved without temperature control by heaters attached to the target cell and heat exchanger. The horizontal axis represents the time normalized to 0 at the moment the needle valve closed. The temperature of the cell dropped below 1.35 K at 1.3 h after the valve closed. In addition, a clear difference in temperature between the heat exchanger and the target cell does not appear in figure 7(a). Heat transport by the siphon method clearly works well. The histogram in figure 7(b) was obtained by expanding the temperature data for the target cell for 8 h (2-10 h) in figure 7(a) to examine the stability. Although the data in figure 7(a) were trimmed by 1/30 for easier viewing, all data, acquired at 10 s intervals, were used in order to increase the statistical precision of the histogram. The temperature stability of the target cell was estimated to be $\pm$ 0.009 K ($\pm$ 3$\sigma$) by Gaussian fitting. The density fluctuation of 0.12\%, converted from the temperature stability, is acceptable for the E15 experiment.

\begin{figure}[htb]
\centering
\includegraphics[width=0.8 \columnwidth,angle=0]{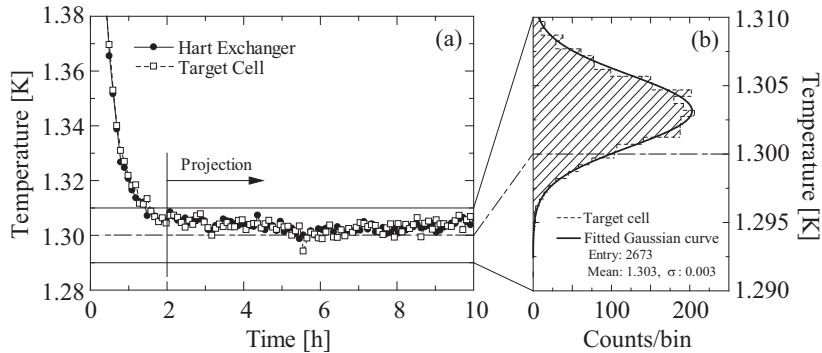}
\caption{Temperature and stability achieved in the heat exchanger and target cell.}
\label{fig.7}       
\end{figure}

Figure 8 shows the correlation between the temperature and the pressure in the heat exchanger and evaporator. Although the values shift slightly compared with those of the International Temperature Scale of 1990 (ITS-90)\cite{its90} owing to the measurement error of the pressure gauge, their behavior agrees well with that of the ITS-90 values. This result strongly suggests that liquid $^{3}$He existed in the heat exchanger.

\begin{figure}[htb]
\centering
\includegraphics[width=0.4 \columnwidth,angle=0]{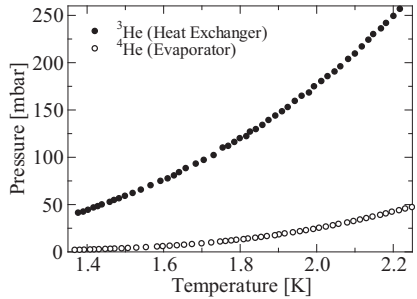}
\caption{Correlation of temperature and pressure in the heat exchanger and evaporator.}
\label{fig.8}       
\end{figure}

Because one month of consecutive operation of the target system is assumed, the consumption of liquid $^{4}$He is very important for describing the direct beam time loss by liquid $^{4}$He dewar exchange. To minimize the consumption of liquid $^{4}$He, the target system is operated in a one-shot mode. Figure 9 shows the time variation in the fluid volume in the evaporator. Because liquid $^{4}$He in the evaporator emptied for about 15 h, the evaporator was refilled with liquid $^{4}$He transported from the outside dewar via separator. At that time, the valve between the tanks and the heat exchanger was closed in order to stop $^{3}$He gas from returning to the tanks owing to the temperature increase in the heat exchanger to 2 K. For the remaining 15 h, the target cell was maintained at a temperature of 1.3 K by isolating the evaporator from the separator. A heat load of 0.21 W was estimated from the decrease in the speed change in the liquid level in figure 9. A total liquid $^{4}$He consumption rate of 50 L/day, including transfer loss and natural evaporation in the dewar, is acceptable for the E15 experiment.

\begin{figure}[htb]
\centering
\includegraphics[width=0.4 \columnwidth,angle=0]{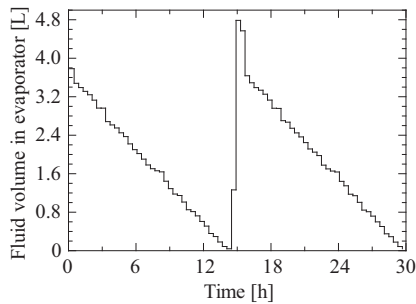}
\caption{Time variation in fluid volume in evaporator.}
\label{fig.9}       
\end{figure}

\begin{table}
\caption{Results of cooling test of Liquid $^{3}$He target.}
\label{SchmidtPL_tab:2} 
\begin{center}
\begin{tabular}{lr|c}
\hline
\multicolumn{2}{c|}{Condition/result}\\ 
\hline
Degree of vacuum                               &     [mbar] &  $<$10$^{-6}$ \\
Leak rate $^{3}$He system $\rightarrow$ vacuum &
                                     [Pa$\cdot$m$^{3}$/sec] & $<$10$^{-10}$ \\
Amount of condensed $^{3}$He gas               &    [liter] &       380/400 \\
Temperature in the target cell                 &        [K] &          1.30 \\
Temperature stability in the target cell       &        [K] &    $\pm$ 0.01 \\
Pressure in the target cell                    &     [mbar] &            33 \\
Heat load to the 1 K parts                     &        [W] &          0.21 \\
Liq. $^{4}$He consumption                      &[liter/day] &            50 \\
\hline
\end{tabular}
\end{center}
\end{table}

\section{Summary}
\label{sec:5}
We successfully developed a liquid $^{3}$He target system for experimental research on deeply bound kaonic nuclear states. In a cooling test, 380 L of $^{3}$He gas were condensed in the cryogenic system. By excellent operation using the siphon method, liquid $^{3}$He was maintained at 1.30 $\pm$ 0.01 K in the target cell for 8 h. The target system is also superior in cost performance owing to its heat load of 0.21 W and liquid $^{4}$He consumption rate of 50 L/day. A radiation shield of thin pure aluminum and a vacuum vessel using thin CFRP cylinders were produced to minimize the amount of material around the target cell. The specially developed target cell made of pure Be and a Be-Al composite satisfied all the requirements; pressure tightness, leak tightness, minimal amount of material, and high x-ray transmission. Experiments using the target system will provide extremely important results in research on kaonic atoms and kaonic nuclei.

\section*{Acknowledgments}
We thank the entire staff at KEK. We are specially, grateful to the members of the KEK cryogenic science center for their support of the performance evaluation test. This developmental research was supported by RIKEN, KEK, Grant-in-Aid for Scientific Research on Priority Areas (17070005), and Grant-in-Aid for Specially Promoted Research (20002003).








\end{document}